\documentclass[12pt]{amsart}
\usepackage[a4paper,margin=1in]{geometry}
\usepackage{amssymb,mathtools}
\usepackage{booktabs}
\usepackage{graphicx}
\graphicspath{{figures/}}
\usepackage{algorithm}
\usepackage{algpseudocode}
\usepackage{hyperref}
\usepackage[capitalise,nameinlink,noabbrev]{cleveref}
\crefname{subsection}{Section}{Sections}
\crefname{algorithm}{Algorithm}{Algorithms}
\crefname{question}{Question}{Questions}

\theoremstyle{plain}
\newtheorem{theorem}{Theorem}[section]
\newtheorem{proposition}[theorem]{Proposition}
\newtheorem{lemma}[theorem]{Lemma}
\newtheorem{corollary}[theorem]{Corollary}
\newtheorem{question}{Question}
\theoremstyle{definition}
\newtheorem{definition}[theorem]{Definition}
\theoremstyle{remark}
\newtheorem{remark}[theorem]{Remark}

\usepackage{xcolor}
\definecolor{titlegrey}{gray}{0.45}

\makeatletter
\def\@settitle{{\centering
  \baselineskip 29pt\relax
  {\Large\bfseries \@title\par}}%
  \vskip 2pt
  {\centering\color{titlegrey}\rule{0.16\textwidth}{0.6pt}\par}%
  \vskip 12pt}
\def\@setauthors{{\centering
  {\large\scshape\authors\par}}%
  \vskip 20pt}
\makeatother

\newcommand{\E}{\mathbb{E}}
\newcommand{\Var}{\operatorname{Var}}
\newcommand{\Cov}{\operatorname{Cov}}
\newcommand{\CV}{\operatorname{CV}}
\newcommand{\AD}{\mathrm{AD}}
\newcommand{\ADh}{\widehat{\mathrm{AD}}}
\newcommand{\Oh}{\widehat{O}}
\newcommand{\Sym}{\mathfrak{S}}
\newcommand{\Varh}{\widehat{\mathrm{Var}}}
\newcommand{\Prob}{\operatorname{Pr}}
\newcommand{\Prt}{\Prob_t}
\newcommand{\Prnu}{\Prob_\nu}

\begin{document}

\title{Bayesian Inference in Partially Observed Critical Systems}
\author{Mihika Singhania}
\date{}

\begin{abstract}
We ask how much of a large random system must be observed to infer a global average with controlled uncertainty. For the random transposition shuffle, a conjugate Bayesian estimator of the average displacement from $m$ sampled cards has relative uncertainty $1/\sqrt{2m}$, independent of the deck size $n$. Because the configuration is a single draw from equilibrium, a floor $R_n$ remains, and no relative tolerance below $\kappa\sqrt{R_n}$ can be certified, even by a census. For the shuffle $R_n=(2n^2+7)/\bigl(5(n-1)^2(n+1)\bigr)$. Before equilibrium, a certificate built on the displacement loses sight after about $\frac n4\ln m$ steps, half as long as the count of fixed cards in the same sample. At the two-dimensional Ising critical point the floor of the energy vanishes, whereas that of the magnetisation stays near $0.072$, with mean-field limit $\sqrt2-1$. For correlated fields the usual 95\% interval can cover the ensemble mean in only 7\% of cases, and no bound on the floor from one realisation holds under stationarity alone.
\end{abstract}

\maketitle
\vspace{-1\baselineskip}

\section{Introduction}

Many large stochastic systems evolve through repeated local interactions while an observer sees only a fraction of their state. The questions of interest are nevertheless global. Has the system equilibrated, and what is the value of some aggregate statistic of its configuration? The classical theory of Markov chains and their mixing times \cite{LevinPeresWilmer2009} presumes access to the full state. When only a subsample is available, equilibrium is judged heuristically, without a finite-sample account of the observation. This is the situation in Markov chain Monte Carlo diagnostics \cite{GelmanRubin1992}.

We invert the question. Suppose the system has run past its mixing time, how many components must be observed before a global equilibrium statistic can be inferred with controlled uncertainty? The question joins the equilibrium theory of finite Markov chains, which supplies a known ground truth, with finite-population sampling theory \cite{Cochran1977}. Our testbed is the random transposition shuffle \cite{DiaconisShahshahani1981}, whose uniform stationary law lets inferential statements be checked against a known truth. The observable is the average displacement $\AD(\pi)=n^{-1}\sum_i|\pi(i)-i|$, a scaled version of Spearman's footrule \cite{DiaconisGraham1977}.

Our main contribution is an account of two sources of error in partial observation. From a uniform sample of $m$ cards, a conjugate Bayesian estimator has relative uncertainty $1/\sqrt{2m}$, independent of $n$. The law of total variance then isolates a second error, the floor, which bounds the tolerance that any sample can certify. Because the relative floor is a generalised susceptibility divided by $nO_\infty^2$, whether a tolerance can be certified depends on how fluctuations scale with system size, a question that statistical physics answers at criticality through finite-size scaling. We evaluate the floor exactly for the shuffle and show that the Bayesian posterior remains valid, and slightly conservative, for the equilibrium value. Before equilibrium, the shuffle can still be followed exactly. The mean displacement after $t$ steps is $\bigl(1-(1-2/n)^t\bigr)\AD_\infty$, so the relative mixing state becomes a parameter with a known law, and a certificate built on the sample mean loses sight after about $\frac n4\ln m$ transpositions. The fixed cards of the same sample detect the shuffle until about $\frac n2\ln m$, twice as long. At the critical point of the two-dimensional Ising model, the floor of the energy vanishes, but that of the magnetisation does not, and the same holds in mean field. For correlated random fields the usual interval loses coverage at the rate the floor predicts, and for a single realisation we prove that no bound on the floor holds under stationarity alone. These results determine when further observation can reduce uncertainty in settings where a global average is judged from part of one configuration, including Markov chain Monte Carlo summaries, environmental sensor networks and the monitoring of large distributed systems.

\cref{sec:prelim} introduces the shuffle, the observable, and the Bayesian formulation, and \cref{sec:methods} describes the computations. \cref{sec:scale} derives the sample-size law and \cref{sec:floor} the floor. \cref{sec:before} follows the shuffle before equilibrium. Sections~\ref{sec:ising}--\ref{sec:estimation} apply the floor to criticality, to correlated fields and to the problem of estimating it from one realisation. \cref{sec:discussion} discusses the assumptions and \cref{sec:conclusion} concludes.

\section{Preliminaries}\label{sec:prelim}

\subsection{The random transposition shuffle}

Let $[n]=\{1,\dots,n\}$ and let $\Sym_n$ denote the symmetric group on $[n]$. A deck of $n$ cards starts in order, and $\pi_t\in\Sym_n$ denotes its state after $t$ transpositions, $\pi_t(i)$ being the position of the card that started at $i$. Throughout, $m$ is the number of observed components, $\nu$ the stationary law and $\mu$ the latent parameter of the Bayesian analysis.

\begin{definition}[Random transposition shuffle]\label{def:shuffle}
At each step, two positions $a_t,b_t\in[n]$ are drawn independently and uniformly, and the cards at those positions are exchanged, so that $\pi_{t+1}=\tau_t\circ\pi_t$ with $\tau_t=(a_t\,b_t)$, the identity when $a_t=b_t$.
\end{definition}

The chain is irreducible, since transpositions generate $\Sym_n$, and aperiodic, since the choice $a_t=b_t$ gives every state a self-loop. Its transition matrix is symmetric, so the uniform law $\nu(\sigma)=1/n!$ is its unique limit \cite{LevinPeresWilmer2009}. Diaconis and Shahshahani~\cite{DiaconisShahshahani1981} showed that the chain mixes with a cutoff at $\tfrac12 n\ln n$ steps. We take $t_{\mathrm{mix}}=\lfloor\tfrac12 n\ln n\rceil$, the nearest integer.

\subsection{Average displacement}

\begin{definition}[Average displacement]\label{def:ad}
For $\sigma\in\Sym_n$ the displacement of card $i$ is $D_i=|\sigma(i)-i|$, and $\AD(\sigma)=\frac1n\sum_{i=1}^n D_i$. Up to the factor $1/n$ this is Spearman's footrule $F=\sum_i D_i$ \cite{DiaconisGraham1977}.
\end{definition}

\begin{proposition}\label{prop:mean}
If $\sigma$ is uniform on $\Sym_n$, then
\begin{equation}\label{eq:mean}
\AD_\infty:=\E[\AD(\sigma)]=\frac{n^2-1}{3n},\qquad
\E|\sigma(i)-i|=\frac{c_i}{n},\qquad
c_i=\frac{(i-1)i}{2}+\frac{(n-i)(n-i+1)}{2}.
\end{equation}
\end{proposition}

\begin{proof}
For each $i$ the image $\sigma(i)$ is uniform on $[n]$, so $\E|\sigma(i)-i|=\frac1n\sum_j|j-i|=c_i/n$. The substitution $i\mapsto n+1-i$ shows that the two terms of $c_i$ have equal sums over $i$, so $\AD_\infty=\frac{1}{n^2}\sum_i(i^2-i)=(n^2-1)/(3n)$.
\end{proof}

The positional mean $c_i/n$ is about $n/4$ at the centre of the deck and $n/2$ at its ends. Any fixed window of positions is therefore biased, and the observer samples uniformly.

\begin{definition}[Random-sample estimator]\label{def:estimator}
For a subset $S\subseteq[n]$ of size $m$ chosen uniformly without replacement, $\ADh_m=\frac1m\sum_{i\in S}|\pi_t(i)-i|$.
\end{definition}

Each index lies in $S$ with probability $m/n$, so $\E_S[\ADh_m]=\AD(\sigma)$ for every $\sigma$ \cite{Cochran1977}, and at equilibrium $\E[\ADh_m]=\AD_\infty$ for every $m$. The inferential difficulty therefore lies entirely in the variance.

\subsection{Bayesian framework}

Let $\mu$ denote the latent equilibrium displacement. Bayes' theorem updates a prior $p(\mu)$ to the posterior $p(\mu\mid\ADh_m)\propto p(\ADh_m\mid\mu)\,p(\mu)$, from which point estimates and $(1-\alpha)$ credible intervals are read off \cite{Bishop2006,Robert2007}. The likelihood and prior are specified in \cref{sec:bayes-methods}.

\section{Methods}\label{sec:methods}

\subsection{Computational framework}

All simulations of the shuffle were written in Python~3.13 with NumPy \cite{Harris2020}, SciPy \cite{Virtanen2020}, pandas, and Matplotlib. The deck is stored as an integer array encoding $\pi_t$. Each step applies \cref{def:shuffle} in place, and $\AD(t)$ is evaluated in a single pass. We swept $n=2,4,\dots,4096$, using $n=10,50,\dots,3000$ to check for artefacts of powers of two, and ran every trajectory to $2\,t_{\mathrm{mix}}$. Each configuration was replicated over $R=1000$ independent runs, and error bars show $\pm1$ standard deviation.

\subsection{Bayesian inference}\label{sec:bayes-methods}

After $t_{\mathrm{mix}}$ steps, we reveal $m$ of the $n$ cards, drawn uniformly without replacement with a fresh subset in each replication, for two to eight values of $m$ per deck up to the census $m=n$. The estimator is a mean of bounded contributions drawn without replacement and is therefore approximately normal \cite{Hajek1960}. We adopt the likelihood $\ADh_m\mid\mu\sim\mathcal N(\mu,\sigma_m^2)$ with the conjugate prior $\mu\sim\mathcal N(\mu_0,\sigma_0^2)$, where $\mu_0=\AD_\infty$ and $\sigma_0=100$ at $n=512$. The posterior is normal, with
\begin{equation}\label{eq:posterior}
\frac{1}{\sigma_{\mathrm{post}}^2}=\frac{1}{\sigma_0^2}+\frac{1}{\sigma_m^2},\qquad
\mu_{\mathrm{post}}=\sigma_{\mathrm{post}}^2\Bigl(\frac{\mu_0}{\sigma_0^2}+\frac{\ADh_m}{\sigma_m^2}\Bigr),
\end{equation}
and the $(1-\alpha)$ credible interval is $\mu_{\mathrm{post}}\pm z_{1-\alpha/2}\,\sigma_{\mathrm{post}}$. The results were insensitive to $\sigma_0\in\{50,75,100,150,200\}$ and to truncating the prior at zero.

\subsection{Exact and Monte Carlo computations}

Footrule moments were checked by enumerating $\Sym_n$ for $n\le9$ in exact arithmetic. For the Ising model at its critical temperature, we enumerated all $2^{16}$ configurations at $L=4$, computed the law of the magnetisation at $L=6$ and $8$ with a row transfer matrix, and obtained energy moments at every $L$ from Kaufman's partition function for the finite torus \cite{FerdinandFisher1969,Kaufman1949}. The latter reproduces the enumeration to fifteen significant figures.

Equilibrium Ising configurations were generated in C with the Wolff single-cluster algorithm \cite{Wolff1989}. A configuration was recorded every $k\approx n/\langle C\rangle$ cluster flips, with $k$ fixed for each run from the mean cluster size $\langle C\rangle$ in a warm-up. Recording after a fixed number of flipped spins would instead bias the ensemble towards order, since large clusters are more frequent in ordered configurations. Each size $L\in\{4,6,8,12,16,24,32,48,64,96,128\}$ was run as two chains of $2\times10^5$ measurements, one from an ordered and one from a random start, and their means agreed with $|z|<2$. Intervals are 95\% block-bootstrap intervals, and slopes are least-squares fits of $\log R_n$ on $\log n$ over $L\ge16$. Mean-field quantities are exact finite sums, correlated fields were generated by fast Fourier transforms on the torus, and all computations for Sections~\ref{sec:floor}--\ref{sec:estimation} use fixed seeds.

\section{Scale-invariant inference}\label{sec:scale}

\subsection{Calibration and variance law}

At $t_{\mathrm{mix}}$ the simulated mean displacement agrees with $\AD_\infty$ to within sampling error for $2\le n\le4096$. At $n=512$ it settles onto $\AD_\infty\approx170.7$ by $t_{\mathrm{mix}}=1597$. The exact shortfall at that time, derived in \cref{sec:before}, is $(1-2/n)^{t_{\mathrm{mix}}}\AD_\infty=0.33$, or $0.19\%$. At $n=512$ the sampling distributions of $\ADh_m$ are centred on $\AD_\infty$ to within $0.4$ and close to normal, with skewness below $0.19$ and excess kurtosis below $0.18$ at $m=32$.

\begin{theorem}\label{thm:variance}
At equilibrium, for a uniform sample of $m\ll n$ cards,
\begin{equation}\label{eq:variance}
\Var(\ADh_m)=\frac{n^2}{18\,m}\bigl(1+O(m/n)\bigr).
\end{equation}
\end{theorem}

\begin{proof}
A single displacement is distributed as $|X-Y|$ for independent $X,Y$ uniform on $[n]$, so $\E D^2=(n^2-1)/6$ and $\Var D=\frac{n^2}{18}\bigl(1+O(n^{-2})\bigr)$. Sampling without replacement multiplies $\Var D/m$ by $(n-m)/(n-1)=1+O(m/n)$, up to a term of order $n$ that is treated exactly in \cref{sec:floor}.
\end{proof}

For a normal likelihood with known variance, the Fisher information about $\mu$ is $I(\mu)=\sigma_m^{-2}=18m/n^2$ \cite{LehmannCasella1998}. The informative quantity is the relative mixing state $\phi=\mu/\AD_\infty$, for which the information is multiplied by $\AD_\infty^2\approx n^2/9$.

\begin{theorem}[Scale invariance]\label{thm:scale}
At equilibrium the relative Fisher information of a uniform sample is $I(\phi)=2m\bigl(1+O(n^{-2})\bigr)$, and
\begin{equation}\label{eq:cv}
\CV(\ADh_m)=\frac{\sqrt{\Var(\ADh_m)}}{\AD_\infty}=\frac{1}{\sqrt{2m}}\bigl(1+O(m/n)\bigr),
\end{equation}
both independent of $n$ to leading order.
\end{theorem}

\begin{proof}
Divide \eqref{eq:variance} by $\AD_\infty^2=\frac{n^2}{9}\bigl(1+O(n^{-2})\bigr)$. In the normal model with $\sigma_m^2=n^2/(18m)$ the information about $\phi$ is $\AD_\infty^2/\sigma_m^2$.
\end{proof}

Within this model, $1/\sqrt{2m}$ is the Cram\'er--Rao bound for $\phi$, and the sample mean attains it. \cref{rem:info} compares this with the exact law of a card before equilibrium.

\subsection{Minimum informative sample}

Requiring the posterior to place probability $1-\alpha$ within $\pm\varepsilon\AD_\infty$ of $\AD_\infty$, that is $z_{1-\alpha/2}\CV\le\varepsilon$, gives
\begin{equation}\label{eq:mstar}
m^\star(\varepsilon)\approx\frac{z_{1-\alpha/2}^2}{2\varepsilon^2},
\end{equation}
about $48$, $192$ and $768$ cards for precision of $20$, $10$ and $5$ per cent at 95\% posterior probability, whatever the size of the deck, provided $m^\star\ll n$. \cref{fig:mstar}(b) tests this law. The measured thresholds are flat in $n$. Among constant, power-law, logarithmic and linear fits to $m^\star(n)$, the constant model minimises both AIC and BIC ($R^2\le0.08$ for the alternatives). The constant that survives is the squared signal-to-noise ratio of one card, $(\E D)^2/\Var D\to2$. The argument applies to any system with bounded contributions and a per-component signal-to-noise ratio bounded away from $0$ and $\infty$, such as particle occupancies \cite{Liggett1985} and server loads \cite{Mitzenmacher2001}.

\section{The equilibrium floor}\label{sec:floor}

\subsection{Sources of error}

The observed configuration is a single draw from $\nu$, so we now work in a general setting. A system has $n$ components, stationary law $\nu$ and real contributions $D_i(x)$ in configuration $x$. The observable is $O(x)=\frac1n\sum_i D_i(x)$, with equilibrium mean $O_\infty=\E_\nu[O]\ne0$ measured from a natural origin, and the observer reports $\Oh_m=\frac1m\sum_{i\in S}D_i(x)$. We write $V_x=\frac1n\sum_i\bigl(D_i(x)-O(x)\bigr)^2$ and $S_x^2=nV_x/(n-1)$.

\begin{lemma}[Exact decomposition]\label{lem:decomp}
For every $1\le m\le n$,
\begin{equation}\label{eq:decomp}
\Var(\Oh_m)=\underbrace{\frac{n-m}{m(n-1)}\,\E_\nu[V_x]}_{\text{sampling term}}+\underbrace{\Var_\nu(O)}_{\text{floor}}.
\end{equation}
\end{lemma}

\begin{proof}
Given $x$, sampling without replacement gives $\E_S[\Oh_m\mid x]=O(x)$ and $\Var_S(\Oh_m\mid x)=V_x\frac{n-m}{m(n-1)}$ \cite{Cochran1977}. The law of total variance yields the first term on averaging over $x$, and the second as $\Var_\nu(O)$.
\end{proof}

For the shuffle, \cref{thm:variance} keeps only the first term, which is appropriate when the target is the deck in hand. When the target is the equilibrium value, the second term is present at every $m$, including the census. Writing $\rho_n^2=O_\infty^2/\E_\nu[V_x]$ for the per-component signal-to-noise ratio, we have
\begin{equation}\label{eq:cv2}
\CV^2(\Oh_m)=\frac{n-m}{m(n-1)}\cdot\frac{1}{\rho_n^2}+R_n,\qquad R_n:=\frac{\Var_\nu(O)}{O_\infty^2}.
\end{equation}
We call $R_n$ the relative floor. It depends on the origin of $O$, and it is the quantity that governs relative tolerances.

\subsection{Certification bound}

To certify that a system has reached equilibrium to relative tolerance $\delta$, the observer tests a mean of $O_\infty$ against $(1-\delta)O_\infty$ at level $\alpha$ with power $1-\beta$. Under the normal approximation this requires
\begin{equation}\label{eq:cert}
\kappa^2\,\frac{\Var(\Oh_m)}{O_\infty^2}\le\delta^2,\qquad\kappa=z_{1-\alpha}+z_{1-\beta},
\end{equation}
where $\kappa\approx2.486$ for $\alpha=0.05$ and $1-\beta=0.80$. Equation~\eqref{eq:mstar} is the same criterion with $z_{1-\alpha/2}$ in place of $\kappa$ and without the floor.

\begin{corollary}[Floor bound]\label{cor:floor}
For every $m$, $\CV^2(\Oh_m)\ge R_n$, and no tolerance $\delta<\kappa\sqrt{R_n}$ can be certified by \eqref{eq:cert} at any sample size. Suppose $|O_\infty|$ is bounded away from zero and infinity on the scale of the contributions, and write $\chi_n=n\Var_\nu(O)$. If $\chi_n=\Theta(1)$, then $R_n=\Theta(1/n)$ and every tolerance becomes certifiable as $n$ grows. If $\chi_n=\Theta(n)$, then $R_n=\Theta(1)$ and tolerances below $\kappa\sqrt{R_n}$ are unreachable at every $m$ and $n$. For $\delta>\kappa\sqrt{R_n}$ the smallest sufficient sample is
\begin{equation}\label{eq:mstar-n}
m_n^\star(\delta)=\biggl\lceil\frac{n}{1+(n-1)\,\rho_n^2\bigl(\delta^2/\kappa^2-R_n\bigr)}\biggr\rceil,
\end{equation}
which rises to $n$ as $\delta\downarrow\kappa\sqrt{R_n}$.
\end{corollary}

\begin{proof}
The sampling term is non-negative, and $R_n=\chi_n/(nO_\infty^2)$. Substituting \eqref{eq:decomp} into \eqref{eq:cert} and solving for $m$ gives \eqref{eq:mstar-n}, which is Cochran's finite-population correction of the limiting value.
\end{proof}

\begin{figure}[t]
\centering
\includegraphics[width=\textwidth]{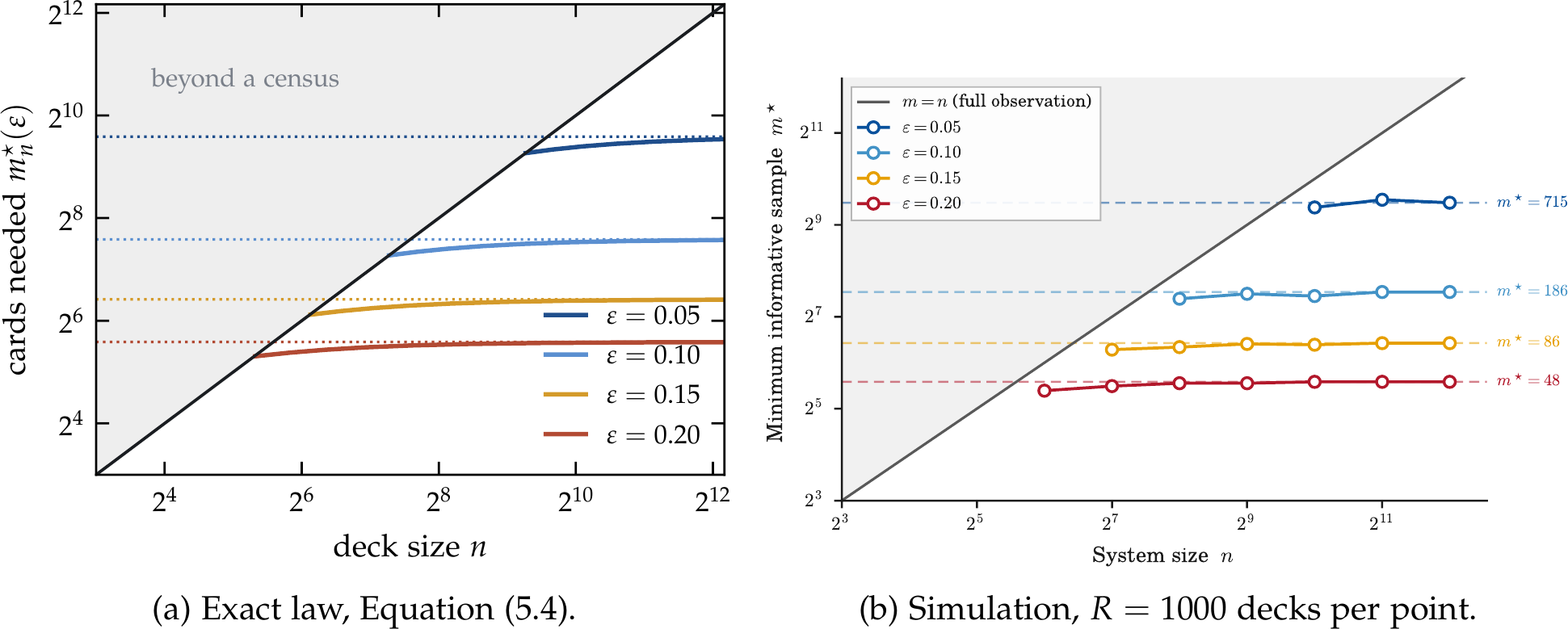}
\caption{Minimum informative sample for relative precision $\varepsilon$ at 95\% posterior probability. (a)~Exact law, \eqref{eq:mstar-n}. (b)~Simulation, $R=1000$ decks per point. Dotted lines in (a) are the leading-order values of \eqref{eq:mstar}. Each exact curve starts on the diagonal $m=n$ and rises to its plateau, which is why the measured $715$ at $\varepsilon=5\%$ lies below $768$.}
\label{fig:mstar}
\end{figure}

\cref{fig:mstar}(a) evaluates \eqref{eq:mstar-n} for the shuffle with $z_{0.975}$ in place of $\kappa$. At $\varepsilon=5\%$ it gives $669$ to $741$ cards over $1024\le n\le4096$, and $715$ at $n=2048$.

\begin{remark}\label{rem:subext}
If $\chi_n\asymp n^a$ with $0<a<1$ and $|O_\infty|$ is bounded away from zero and infinity, then $R_n\asymp n^{a-1}\to0$. In that regime the open question is the normal approximation for correlated contributions, not the floor. When $O_\infty\to0$, the exponent alone does not decide (\cref{sec:ising}).
\end{remark}

\subsection{The exact floor of the shuffle}

Normalising the contributions to $D_i/n\in[0,1]$ leaves the relative floor unchanged. This is the scale on which $\AD_\infty/n\to1/3$ and on which $\chi_n$ is stated. The quantities $\E_\nu[V_x]$ and $\Delta_n$ are stated in positions. The variance of the footrule is classical \cite{DiaconisGraham1977}.

\begin{proposition}\label{prop:floor}
For $\sigma$ uniform on $\Sym_n$, $\Var F=(n+1)(2n^2+7)/45$. Consequently
\begin{equation}\label{eq:Rn}
R_n=\frac{\Var F}{(\E F)^2}=\frac{2n^2+7}{5(n-1)^2(n+1)},\qquad\lim_{n\to\infty}nR_n=\frac25,
\end{equation}
and $\E_\nu[V_x]=(n-2)(n+1)(5n^2+n+12)/(90n^2)$, so that $\rho_n^2\to2$ and $\chi_n=n\Var_\nu(\AD/n)\to2/45$.
\end{proposition}

\begin{proof}
For $i\ne j$ the pair $(\sigma(i),\sigma(j))$ is uniform on ordered pairs of distinct positions, so $\E[D_iD_j]=\bigl(c_ic_j-\sum_a|a-i||a-j|\bigr)/\bigl(n(n-1)\bigr)$ and $\E[D_i^2]=\frac1n\sum_a(a-i)^2$. With $\Sigma_2=\sum_{a,i}(a-i)^2=n^2(n^2-1)/6$, $\sum_ic_i=n(n^2-1)/3$ and $\sum_ic_i^2=n(n^2-1)(7n^2-8)/60$,
\[
\E F^2=\frac{\Sigma_2}{n}+\frac{\bigl(\sum_ic_i\bigr)^2-2\sum_ic_i^2+\Sigma_2}{n(n-1)},
\]
and subtracting $(\E F)^2=(n^2-1)^2/9$ gives $\Var F$. Finally $\E_\nu[V_x]=\frac{n^2-1}{6}-\AD_\infty^2-\Var_\nu(\AD)$.
\end{proof}

\begin{figure}[t]
\centering
\includegraphics[width=\textwidth]{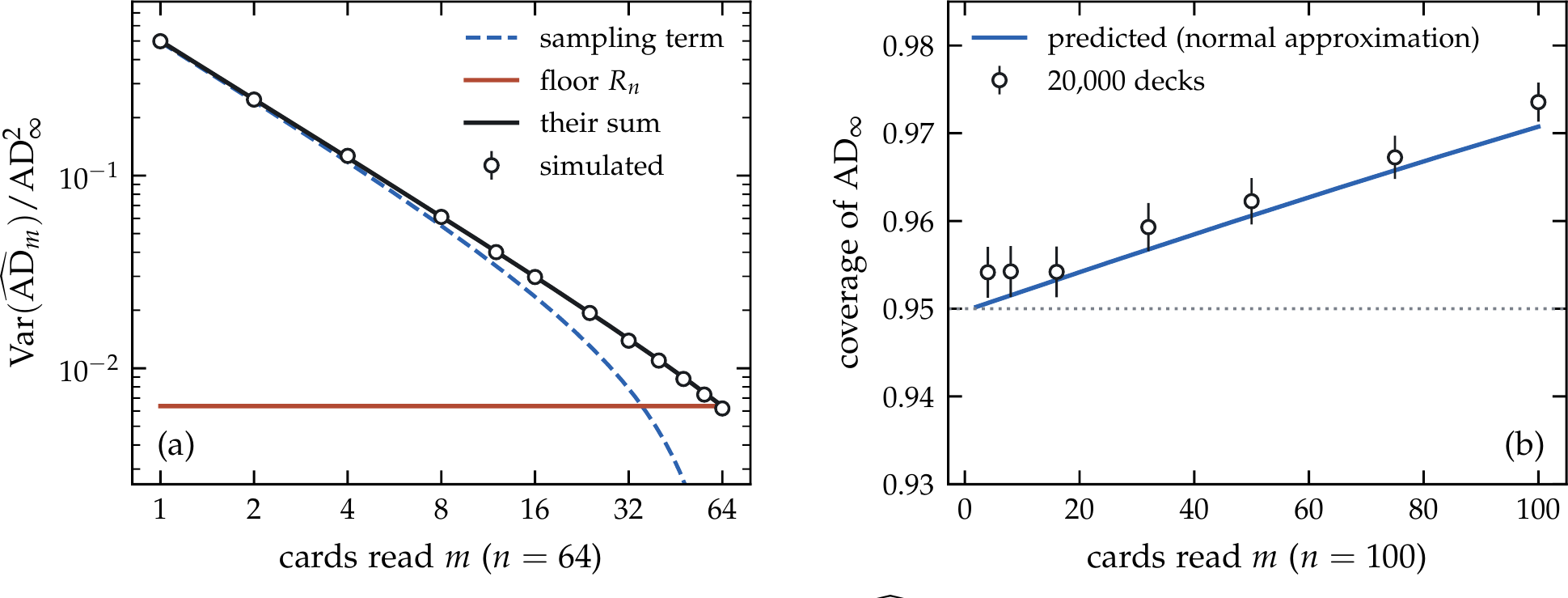}
\caption{The floor of the shuffle. (a)~Relative variance of $\ADh_m$ at $n=64$ over $20{,}000$ uniform decks, with 95\% bootstrap intervals, against the two exact terms of \cref{lem:decomp}. The sampling term vanishes at the census, and the floor does not. Their sum matches the simulation with no fitted parameter. (b)~Coverage of $\AD_\infty$ by the 95\% credible interval of \eqref{eq:posterior}, with $\sigma_m^2=n^2/(18m)$, at $n=100$ over $20{,}000$ decks, against \eqref{eq:coverage}. Since $\Delta_n<0$ in \eqref{eq:Sbar}, the interval is conservative and its coverage rises with $m$.}
\label{fig:shufflefloor}
\end{figure}

Enumeration for $2\le n\le9$ reproduces every quantity in \cref{prop:floor}. For example, $\Var F=20/9$ and $R_3=5/16$ at $n=3$. The relative error $\sqrt{R_n}$ is $21.6\%$ at $n=10$, $2.0\%$ at $n=10^3$ and $0.063\%$ at $n=10^6$. A ten-card deck therefore retains a relative error of about a fifth even under a census, while a large deck can be certified to $\delta=10\%$ with $m_n^\star\to\kappa^2/(2\delta^2)\approx309$ cards. \cref{fig:shufflefloor}(a) shows both terms of \cref{lem:decomp} on simulated decks.

\subsection{Posterior calibration}

The posterior \eqref{eq:posterior} carries only the sampling variance, so it describes the configuration in hand. For the equilibrium value $O_\infty$, the hierarchical model $x\sim\nu$, $\Oh_m\mid x$ has the variance \eqref{eq:decomp}, and calibration depends on the sign of a single quantity.

\begin{proposition}\label{prop:calib}
Let $\bar S_n^2=\E_\nu[S_x^2]$ and $\theta_i=\E_\nu[D_i]$, and write
\[
H_n=\frac1n\sum_i(\theta_i-O_\infty)^2,\qquad\bar c_n=\frac{1}{n(n-1)}\sum_{i\ne j}\Cov_\nu(D_i,D_j).
\]
For every $1\le m\le n$,
\begin{equation}\label{eq:Sbar}
\Var(\Oh_m)=\frac{\bar S_n^2}{m}+\Delta_n,\qquad\Delta_n:=\Var_\nu(O)-\frac{\bar S_n^2}{n}=\bar c_n-\frac{H_n}{n-1}.
\end{equation}
Under the normal approximation the interval $\Oh_m\pm z_{1-\alpha/2}\,\bar S_n/\sqrt m$ covers $O_\infty$ with probability
\begin{equation}\label{eq:coverage}
2\,\Phi\Bigl(z_{1-\alpha/2}\bigl(1+m\Delta_n/\bar S_n^2\bigr)^{-1/2}\Bigr)-1,
\end{equation}
which is conservative for $\Delta_n<0$, exact for $\Delta_n=0$, and decreasing in $m$ for $\Delta_n>0$. With the floor in the likelihood, the relative information is $I_n(\phi)=O_\infty^2/\bigl(\bar S_n^2/m+\Delta_n\bigr)$, equal to $1/R_n$ at the census. For the shuffle,
\begin{equation}\label{eq:Delta}
\Delta_n=-\frac{(n+1)(n^3-5n^2-4n-10)}{90\,n^2(n-1)}<0\qquad(n\ge6).
\end{equation}
\end{proposition}

\begin{proof}
Substituting $\E_\nu[V_x]=(n-1)\bar S_n^2/n$ into \eqref{eq:decomp} gives the first identity, and expanding $\Var_\nu(O)$ and $\E_\nu\sum_i(D_i-O)^2$ in the moments of the $D_i$ gives the second. The coverage follows because the half-width uses $\bar S_n^2/m$. The value for the shuffle follows from \cref{prop:floor}.
\end{proof}

For the shuffle $\Delta_n\approx-n/90$, and the negative covariance of displacements and the spread of positional means contribute equally to it. The posterior \eqref{eq:posterior}, whose $\sigma_m^2=n^2/(18m)$ equals $\bar S_n^2/m$ to leading order, therefore remains valid for $\AD_\infty$ at every $m$, and slightly conservative. In \cref{fig:shufflefloor}(b) its 95\% interval covers $\AD_\infty$ in $95.4\%$ of decks at $m=4$ and in $97.4\%$ at the census, against $95.1\%$ and $97.1\%$ predicted by \eqref{eq:coverage}. For the systems of \cref{sec:ising,sec:fields}, $\Delta_n>0$.

\section{Certification before equilibrium}\label{sec:before}

The certificate of \cref{cor:floor} is meant for a system that may not have mixed, yet so far every deck was treated as a draw from $\nu$. The shuffle can be followed exactly before equilibrium, so we can ask how long a certificate keeps its power. Function-specific mixing times \cite{Rabinovich2020} start from the same observation, that a statistic can settle before the chain does. We write $\Prt$, $\E_t$ and $\Var_t$ for the law of $\pi_t$ started from the identity, and set $\lambda=1-2/n$.

\subsection{The mixing state as a parameter}

\begin{proposition}\label{prop:exactmean}
For every $n$, $t$ and $i,j\in[n]$,
\begin{equation}\label{eq:onecard}
\Prt\bigl(\pi_t(i)=j\bigr)=\lambda^t\,\mathbf 1\{j=i\}+\frac{1-\lambda^t}{n}.
\end{equation}
Consequently, for every $1\le m\le n$,
\begin{equation}\label{eq:phit}
\E_t[\ADh_m]=\E_t[\AD(\pi_t)]=\phi(t)\,\AD_\infty,\qquad\phi(t)=1-\Bigl(1-\frac2n\Bigr)^t,
\end{equation}
and $\E[\AD(\pi_{t+1})\mid\pi_t]=\lambda\,\AD(\pi_t)+(1-\lambda)\AD_\infty$.
\end{proposition}

\begin{proof}
A card moves only if exactly one of $a_t,b_t$ equals its position, which has probability $q=2(n-1)/n^2$, and it then moves to a uniform position among the other $n-1$. Its position is therefore a Markov chain with transition matrix $(1-q)I+q(J-I)/(n-1)$, where $J$ is the matrix of ones. This matrix fixes constant vectors and multiplies every vector with zero sum by $1-qn/(n-1)=\lambda$, which gives \eqref{eq:onecard}. The displacement vanishes at $j=i$, so averaging $|j-i|$ against \eqref{eq:onecard} gives $(1-\lambda^t)c_i/n$, and summing over $i$ gives \eqref{eq:phit}. The last identity is the case of one step, applied to every card given $\pi_t$.
\end{proof}

The relative mixing state $\phi=\mu/\AD_\infty$ of \cref{sec:scale} is therefore an exact function of time. At $n=512$ and $t_{\mathrm{mix}}=1597$ it falls short of $1$ by $\lambda^{t_{\mathrm{mix}}}=1.9\times10^{-3}$, a bias of $0.33$ in $\AD_\infty\approx170.7$ and about a fifteenth of the census error $\sqrt{R_n}\,\AD_\infty$. The same formula gives $1+(n-1)\lambda^t$ expected fixed points, so at $t_{\mathrm{mix}}$ the deck still has about two of them, against one at equilibrium.

\begin{lemma}\label{lem:vart}
For every $n$, $t$ and $1\le m\le n$,
\[
\Var_t\bigl(\AD(\pi_t)\bigr)\le\frac{n^2}{6(n-1)},\qquad
\Var_t(\ADh_m)\le\frac{n-m}{m(n-1)}\,\frac{n^2}{16}+\frac{n^2}{6(n-1)}.
\]
\end{lemma}

\begin{proof}
By \cref{prop:exactmean}, $Y_s=\lambda^{-s}\bigl(\AD(\pi_s)-\AD_\infty\bigr)$ is a martingale, so
\[
\Var_t\bigl(\AD(\pi_t)\bigr)=\sum_{s<t}\lambda^{2(t-s-1)}\,\E\Var\bigl(\AD(\pi_{s+1})\mid\pi_s\bigr).
\]
A step exchanges the cards at $a_s$ and $b_s$, whose displacements change by at most $|a_s-b_s|$ each, so the conditional variance is at most $4\E|a_s-b_s|^2/n^2<2/3$. Summing over $s$ with $1-\lambda^2=4(n-1)/n^2$ gives the first bound. The proof of \cref{lem:decomp} uses only the law of total variance, so \eqref{eq:decomp} holds under $\Prt$ as well. By \eqref{eq:onecard}, $\E_t[V_x]\le(1-\lambda^t)(n^2-1)/6-(1-\lambda^t)^2\AD_\infty^2$, which is at most $n^2/16$ for every value of $\lambda^t$.
\end{proof}

\subsection{Half the window}

Two statistics of the same sample can now be compared, the sample mean $\ADh_m$ on which \eqref{eq:cert} is built, and the number $K$ of sampled cards $i\in S$ with $\pi_t(i)=i$. Write $\sigma_m^2=\Var_\nu(\ADh_m)$, which \cref{lem:decomp,prop:floor} give in closed form.

\begin{theorem}[Half window]\label{thm:halfwindow}
Let $n\to\infty$, let $m=m_n\to\infty$ with $m\le n$, let $t=t_n$, and fix $\alpha\in(0,1)$.
\begin{enumerate}
\item[(a)] If $\lambda^t\sqrt m\to\infty$, the test that rejects equilibrium when $\ADh_m<\AD_\infty-\sigma_m/\sqrt\alpha$ has level at most $\alpha$ and power tending to one.
\item[(b)] If $m/n\to0$ and $\lambda^t\sqrt m\to0$, then $\sup_x\bigl|\Prt(\ADh_m\le x)-\Prnu(\ADh_m\le x)\bigr|\to0$, so every test that rejects when $\ADh_m$ crosses a threshold has power tending to its level.
\item[(c)] If $m\lambda^t\to\infty$, the test that rejects equilibrium when $K\ge\lceil1/\alpha\rceil$ has level at most $\alpha$ and power tending to one.
\end{enumerate}
\end{theorem}

Since $\lambda^t=\exp\bigl(-2t/n+O(t/n^2)\bigr)$, parts (a) and (b) compare $t$ with $\frac n4\ln m$ and part (c) compares it with $\frac n2\ln m$, in each case on the scale $n$ of the cutoff window. Built on the sample mean, the certificate loses sight at about $\frac n4\ln m$ steps, while the fixed cards of the same sample detect the shuffle until about $\frac n2\ln m$, twice as long. For the census, (a) gives detection while $\lambda^t/\sqrt{R_n}\to\infty$, that is, up to about $\frac n4\ln n$ steps, half of the cutoff time.

\begin{proof}
For (a), Chebyshev's inequality gives the level. \cref{lem:vart}, and its limit as $t\to\infty$, bound both $\Var_t(\ADh_m)$ and $\sigma_m^2$ by $n^2/(2m)$, using $1/(n-1)\le2/m$. Under $\Prt$ the mean shifts down by $\lambda^t\AD_\infty\ge\lambda^tn/4$, which exceeds $\sigma_m/\sqrt\alpha$ by a factor of order $\lambda^t\sqrt m\to\infty$, and Chebyshev's inequality under $\Prt$ gives the power.

For (b), write $(\ADh_m-\AD_\infty)\sqrt{18m}/n=T_1+T_2$, where $T_2=(\AD(\pi_t)-\AD_\infty)\sqrt{18m}/n$ satisfies $\E_t|T_2|\le\lambda^t\sqrt{2m}+\sqrt{3m/(n-1)}\to0$. Given $\pi_t$, $T_1$ is a scaled mean of $m$ draws without replacement from the displacements of the deck, which are bounded by $n$. Their variance $V_{\pi_t}$ has mean $\frac{n^2}{18}(1+o(1))$ because $\lambda^t\to0$, and the martingale argument of \cref{lem:vart}, applied to $\frac1n\sum_iD_i^2$, together with $\Var_t(\AD^2)\le4n^2\Var_t(\AD)$, shows that its variance is $O(n^3)$. So $V_{\pi_t}/n^2\to1/18$ in probability, and H\'ajek's theorem \cite{Hajek1960} makes $T_1$ asymptotically standard normal given $\pi_t$. The same holds under $\nu$, and P\'olya's theorem turns the two normal limits into uniform convergence of the distribution functions.

For (c), $\E_\nu K=m/n\le1$, so Markov's inequality gives the level. Under $\Prt$, let $U_i$ be the event that position $i$ is drawn neither as $a_s$ nor as $b_s$ in the first $t$ steps. On $U_i$ card $i$ never moves, so $K\ge N=\sum_{i\in S}\mathbf 1_{U_i}$. Since $\Prob(U_i)=(1-1/n)^{2t}\ge\lambda^t$ and $\Prob(U_i\cap U_j)=(1-2/n)^{2t}\le\Prob(U_i)\Prob(U_j)$ for $i\ne j$, $N$ has mean at least $m\lambda^t$ and variance at most its mean, and Chebyshev's inequality gives $\Prt(N<m\lambda^t/2)\le4/(m\lambda^t)\to0$.
\end{proof}

No statistic of the sample can be expected to push (c) much further. The positions of the cards in $S$ form the exclusion process of $m$ labelled particles on the complete graph, whose $\frac12$-lazy version has a cutoff at $n\ln m+O(n)$ steps \cite{LacoinLeblond2011,PatkoPete2019}. Since the lazy chain stays put at half of its steps, this corresponds to about $\frac n2\ln m$ steps of the shuffle, which is where (c) stops.

In the criterion \eqref{eq:cert} the deficit at time $t$ is $\delta=\lambda^t$, so the certificate detects the shuffle with power $1-\beta$ until $t_{\AD}(m)=\ln\bigl(1/(\kappa\CV_m)\bigr)/\ln(1/\lambda)$, with $\CV_m$ from \eqref{eq:cv2}. At $n=512$ this is $353$, $537$ and $681$ steps for $m=48$, $192$ and the census, against $\frac n2\ln m=991$, $1346$ and $1597$. We checked this on $4{,}000$ decks at $n=512$, run from the identity with a fresh sample at each recorded time. At $t=800$, half of $t_{\mathrm{mix}}$, the one-sided $5\%$ test of \eqref{eq:cert} with $m=192$ rejects equilibrium in $22\%$ of decks, while the test $K\ge3$ on the same cards, whose level is $0.7\%$, rejects in $99\%$. At $t=1000$ the rates are $12\%$ and $79\%$, and for the census $19\%$ and $99.8\%$, with $\operatorname{Fix}\ge4$ at level $1.9\%$.

\begin{remark}\label{rem:hoeffding}
\cref{prop:exactmean} extends to every statistic $W_f(\pi)=\sum_if\bigl(i,\pi(i)\bigr)$, for which $\E_tW_f-\E_\nu W_f=\lambda^t\operatorname{tr}d_f$, where $d_f$ is $f$ with its row and column means removed. Hoeffding \cite{Hoeffding1951} showed that $\Var_\nu W_f=\|d_f\|^2/(n-1)$ in the Frobenius norm, and $\operatorname{tr}d_f$ is the inner product of $d_f$ with $I-J/n$, whose norm is $\sqrt{n-1}$. By the Cauchy--Schwarz inequality the standardised deficit of a non-constant $W_f$ is at most $(n-1)\lambda^t$, with equality exactly when $W_f$ is an affine function of the number of fixed points. For the footrule it is $\lambda^t/\sqrt{R_n}\approx\lambda^t\sqrt{5n/2}$. Among all such statistics, the fixed-point count keeps its power longest.
\end{remark}

\begin{remark}\label{rem:info}
One observed card has the law \eqref{eq:onecard} with $\theta=1-\phi=\lambda^t$, and its Fisher information about $\phi$ is $(1-1/n)^2/\bigl(\theta+(1-\theta)/n\bigr)+(n-1)/\bigl(n(1-\theta)\bigr)$, which tends to $n-1$ as $\phi\to1$. The normal model of \cref{sec:bayes-methods} gives the same card information $2$ (\cref{thm:scale}). When the question is whether the deck has mixed, the exact law of the sampled cards therefore carries far more information than the posterior \eqref{eq:posterior} uses, although that posterior remains the right tool for the displacement of the deck in hand.
\end{remark}

\section{Ising criticality}\label{sec:ising}

\subsection{Energy and magnetisation}

Because $\chi_n$ and $O_\infty$ depend on the observable, so does $R_n$. The two-dimensional Ising model at $T_c=2/\ln(1+\sqrt2)$ \cite{Onsager1944}, whose critical exponents are known exactly \cite{Baxter1982,Yang1952}, provides a controlled test. Each site of an $L\times L$ torus carries a spin $s_i=\pm1$, $n=L^2$, and configurations have weight proportional to $\exp\bigl(T^{-1}\sum_{\langle ij\rangle}s_is_j\bigr)$. The energy density $e=\frac1n\sum_iD_i$, with $D_i=-s_i(s_{i\to}+s_{i\downarrow})$, has the form required by \cref{lem:decomp}, and its mean tends to $-\sqrt2$. The magnetisation magnitude $|M|=\bigl|\frac1n\sum_is_i\bigr|$ is not an average of per-site values, so we state its floor at the census, which returns $|M(x)|$ exactly with relative error $\sqrt{R_n}$, where $R_n[|M|]=\Var_\nu(|M|)/\langle|M|\rangle^2$.

\subsection{Results}

\cref{tab:ising} compares the floors of the two observables. The floor of the energy vanishes, falling from $0.103$ at $L=4$ to $3.94\times10^{-4}$ at $L=128$. Every Monte Carlo value lies within $1\%$ of Kaufman's exact value, and the effective slope over $16\le L\le128$ is $-0.864$, the same as for the exact values.

\begin{table}[t]
\centering
\caption{Relative floors at $T_c$ from $4\times10^5$ Wolff measurements per size, with 95\% block-bootstrap intervals for $|M|$. Exact values are from enumeration ($L=4$), the transfer matrix ($L=8$) and Kaufman's partition function. The Binder ratio $\langle M^2\rangle^2/\langle M^4\rangle$ is from Monte Carlo.}
\label{tab:ising}
\small
\setlength{\tabcolsep}{4pt}
\begin{tabular}{rccccc}
\toprule
$L$ & $R_n[e]$, Monte Carlo & $R_n[e]$, exact & $R_n[|M|]$, Monte Carlo & $R_n[|M|]$, exact & Binder ratio\\
\midrule
4   & 0.1030 & 0.1028 & 0.0692 [0.0685, 0.0698] & 0.0692 & 0.8707\\
8   & 0.0412 & 0.0414 & 0.0701 [0.0694, 0.0708] & 0.0706 & 0.8620\\
16  & 0.0142 & 0.0143 & 0.0712 [0.0704, 0.0720] & -- & 0.8586\\
32  & $4.519\times10^{-3}$ & $4.518\times10^{-3}$ & 0.0717 [0.0710, 0.0725] & -- & 0.8571\\
64  & $1.363\times10^{-3}$ & $1.359\times10^{-3}$ & 0.0721 [0.0715, 0.0728] & -- & 0.8562\\
128 & $3.940\times10^{-4}$ & $3.958\times10^{-4}$ & 0.0720 [0.0714, 0.0727] & -- & 0.8565\\
\bottomrule
\end{tabular}
\end{table}

The floor of the magnetisation does not vanish. $R_n[|M|]$ is flat, with slope $+0.002$ against $\log n$ and mean $0.0718$ over $L\ge16$, which is a relative error $\sqrt{R_n}\approx26.8\%$ at every size tested. It agrees with the exact values for $L\le8$, and the Binder ratio settles at the universal torus value of about $0.8562$ \cite{SalasSokal2000}. A fit of $\chi_n=n\Var_\nu(|M|)\propto n^a$ gives $a=0.877$, consistent with $\gamma/(2\nu)=7/8$. Here $\beta$, $\gamma$ and $\nu$ denote critical exponents. \cref{fig:ising} shows the two floors and the spread of census values behind them.

Finite-size scaling explains this behaviour. Since $R_n=\chi_n/(nO_\infty^2)$ with $\chi_n\propto n^{7/8}$ and $\langle|M|\rangle\propto L^{-\beta/\nu}=n^{-1/16}$, we get $R_n\propto n^0$, and $R_n+1=\langle M^2\rangle/\langle|M|\rangle^2$ tends to a universal ratio, as the Binder ratio does. The susceptibility is sub-extensive ($a<1$), yet the floor persists because $O_\infty\to0$, the case excluded in \cref{rem:subext}. For the energy, the specific heat diverges logarithmically \cite{FerdinandFisher1969,Onsager1944}, so $\chi_n[e]=T_c^2c_L$ grows like $\ln n$ and $R_n[e]\propto(\ln n)/n$. The measured $-0.86$ is an effective exponent. On the same configurations, the energy can be certified to any relative tolerance as $L$ grows, while the order parameter keeps a relative dispersion of $27\%$.

\subsection{Mean-field limit}

In the Curie--Weiss model the weight of a configuration depends only on $M=(2k-N)/N$, with $\Pr(k)\propto\binom Nk\exp\bigl(NM^2/(2T)\bigr)$ and $T_c=1$, so every moment is a finite sum. Exact sums for $10^3\le N\le10^6$ give $\Var(M)\propto N^{a-1}$ with $a=0.0006$ at $T=1.5\,T_c$ and $a=1.0006$ at $T=0.8\,T_c$. Since the signed magnetisation has mean zero, we use $|M|$.

\begin{figure}[t]
\centering
\includegraphics[width=\textwidth]{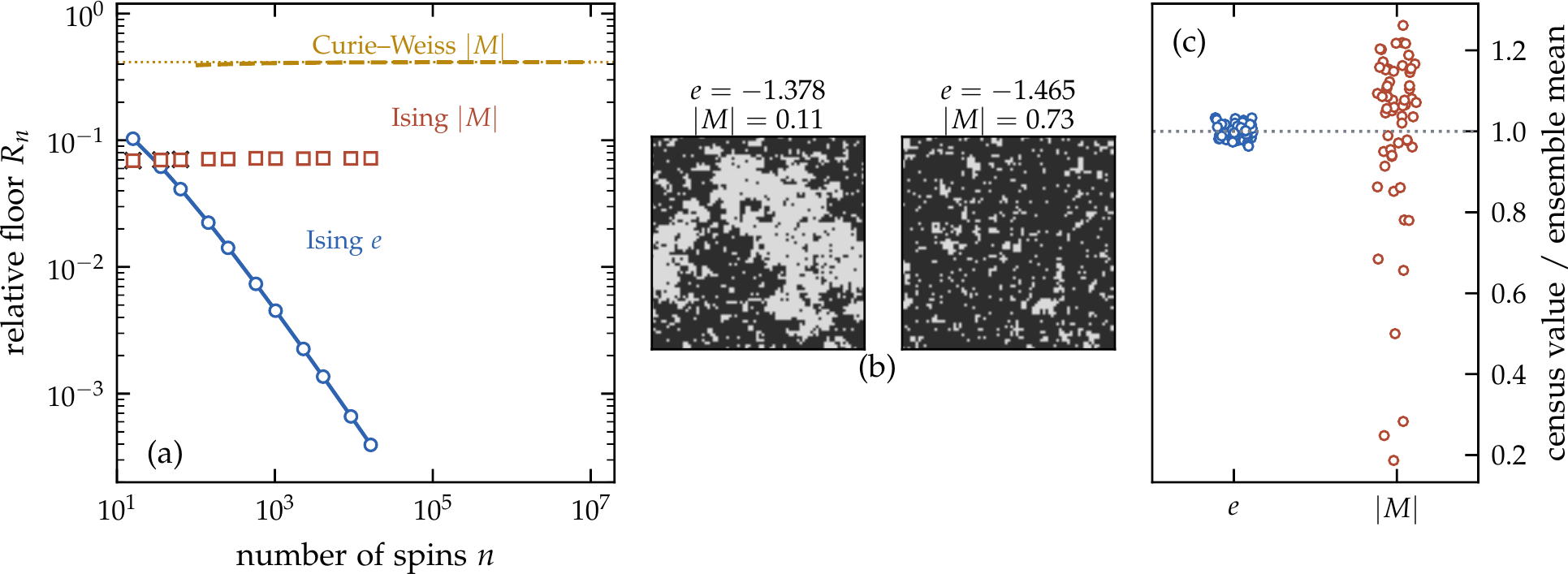}
\caption{Floors at criticality. (a)~Relative floors against the number of spins, for the Ising energy (circles) on Kaufman's exact curve, Ising $|M|$ (squares, with exact values at $L\le8$ as crosses), and the exact Curie--Weiss $|M|$ at $T_c$, which tends to $\sqrt2-1$. (b)~Two equilibrium configurations at $L=128$, dark sites $s_i=+1$, whose energies differ by $6\%$ but whose magnetisations are $0.11$ and $0.73$. (c)~Census values of $60$ successive configurations at $L=128$, divided by the ensemble mean. The energy lies within $4\%$ of its mean, while $|M|$ ranges from $0.2$ to $1.3$ times its mean.}
\label{fig:ising}
\end{figure}

\begin{proposition}\label{prop:cw}
For the Curie--Weiss model, $R_N[|M|]\to0$ for $T<T_c$, $R_N[|M|]\to\sqrt2-1$ at $T=T_c$, and $R_N[|M|]\to\pi/2-1$ for $T>T_c$.
\end{proposition}

\begin{proof}[Proof (sketch)]
At $T_c$, $N^{1/4}M$ converges in law to the density proportional to $e^{-y^4/12}$ \cite{EllisNewman1978}, and we take its first two moments to converge, as the exact sums confirm. Since $\int_0^\infty y^ke^{-y^4/12}\,dy=\frac14\,12^{(k+1)/4}\,\Gamma\bigl(\frac{k+1}{4}\bigr)$, we have $\langle M^2\rangle/\langle|M|\rangle^2\to\Gamma(\frac14)\Gamma(\frac34)/\Gamma(\frac12)^2=\sqrt2$. Above $T_c$, $\sqrt N M$ is asymptotically normal, and the ratio tends to $\pi/2$. Below $T_c$, $|M|$ concentrates at the positive root of $M=\tanh(M/T)$, so $R_N=O(1/N)$.
\end{proof}

Exact sums up to $N=10^7$ agree with all three limits (\cref{fig:ising}a). The non-vanishing critical floor is thus proved in mean field and measured on the lattice. \cref{tab:summary} summarises the observables studied. The fixed-point fraction of the shuffle, whose count has mean and variance $1$, has $R_n=1$ exactly although $\chi_n=1/n$, so even a census cannot certify it below $\kappa$.

\begin{table}[t]
\centering
\caption{Floors of the observables studied, as the system grows.}
\label{tab:summary}
\small
\begin{tabular}{lccc}
\toprule
Observable & $\chi_n$ & $O_\infty$ & $R_n$\\
\midrule
Shuffle, $\AD/n$ & $\to2/45$ & $\to1/3$ & $\sim2/(5n)$\\
Shuffle, fixed-point fraction & $1/n$ & $1/n$ & $=1$\\
Ising $e$ at $T_c$ & $\propto\ln n$ & $\to-\sqrt2$ & $\propto(\ln n)/n$\\
Ising $|M|$ at $T_c$ & $\propto n^{7/8}$ & $\propto n^{-1/16}$ & $\approx0.072$\\
Curie--Weiss $|M|$ at $T_c$ & $\propto N^{1/2}$ & $\propto N^{-1/4}$ & $\to\sqrt2-1$\\
Field mean, $\xi$ fixed & $\Theta(1)$ & $\mu$ & $\propto1/n$\\
Field mean, $\xi\propto L$ & $\Theta(n)$ & $\mu$ & $\Theta(1)$\\
\bottomrule
\end{tabular}
\end{table}

\section{Correlated random fields}\label{sec:fields}

A stationary Gaussian random field is a simple model of a spatially correlated quantity, such as the concentration of a pollutant over a district. On an $L\times L$ torus, standard normal noise $w$ is smoothed by a Gaussian kernel $G$ of width $\xi$ with $\sum G=1$, and $X=\mu+\sigma f$ with $f=(G\circledast w)/\bigl(\sum G^2\bigr)^{1/2}$, so that every site has variance $\sigma^2$. The observable is the spatial mean $O$, read at $m$ random sites, and the target is the ensemble mean $\mu$.

\begin{proposition}\label{prop:field}
On the torus, $\Var_\nu(O)=\sigma^2/\bigl(n\sum_xG(x)^2\bigr)$, so $R_n\approx4\pi\sigma^2\xi^2/(\mu^2L^2)$ for $1\ll\xi\ll L$, and $\Delta_n=\bigl(n\Var_\nu(O)-\sigma^2\bigr)/(n-1)>0$ for $\xi>0$.
\end{proposition}

\begin{proof}
Circular convolution preserves sums, so $\sum_xf(x)=\bigl(\sum_yw(y)\bigr)/\bigl(\sum G^2\bigr)^{1/2}$, with variance $n/\sum G^2$, and $\sum G^2\approx(4\pi\xi^2)^{-1}$. By stationarity $H_n=0$ in \eqref{eq:Sbar}, and $\sum G^2<1$ for a non-degenerate kernel.
\end{proof}

With $\xi$ fixed the floor falls like $1/n$, while with $\xi\propto L$ it is constant, which is the extensive case of \cref{cor:floor}. Since $\Delta_n>0$, \cref{prop:calib} predicts that the usual interval $\bar X\pm1.96\,s/\sqrt m$, which coincides with the weak-prior posterior, fails. We tested it for $L=64$, $m=200$, $\mu=1$ and $\sigma=0.3$, with $2{,}000$ fields at each correlation length.

\begin{figure}[t]
\centering
\includegraphics[width=\textwidth]{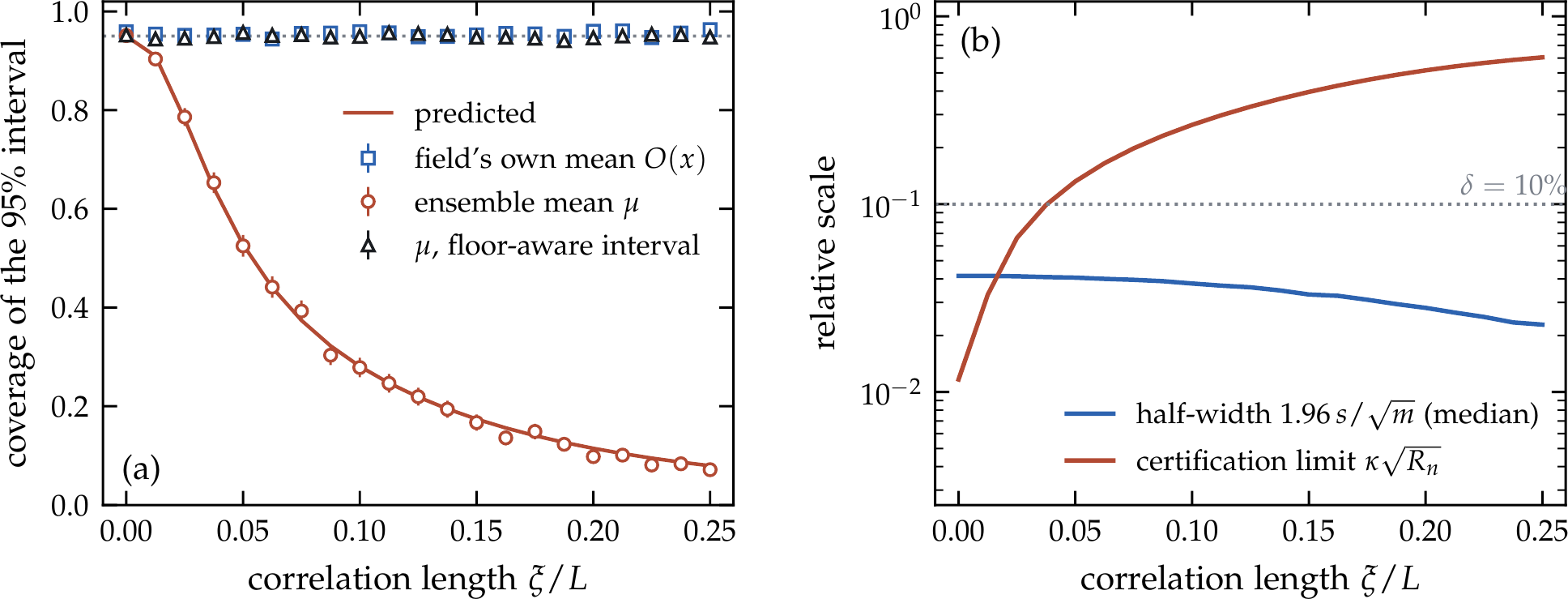}
\caption{A correlated field on a $64\times64$ torus read at $m=200$ random sites, $2{,}000$ fields per point. (a)~Coverage of the usual 95\% interval for the field's own mean and for $\mu$, with the prediction of \eqref{eq:coverage} and the coverage of an interval that includes the floor. The interval remains valid for $O(x)$, but its coverage of $\mu$ collapses onto the predicted curve, while the interval with the floor stays near 95\%. (b)~The median relative half-width of the interval and the certification limit $\kappa\sqrt{R_n}$. As correlations lengthen, the interval narrows while the limit rises, exceeding $\delta=10\%$ near $\xi/L=0.04$.}
\label{fig:field}
\end{figure}

\cref{fig:field}(a) shows that the interval remains valid for the field's own mean, covering it in $94$ to $96\%$ of fields, because the sites are random. Its coverage of $\mu$ falls from $95\%$ for independent sites to $28\%$ at $\xi/L=0.1$ and $7\%$ at $\xi/L=0.25$, following the curve of \eqref{eq:coverage}. Meanwhile its relative half-width shrinks from $0.041$ to $0.023$ (\cref{fig:field}b). The spread within a field estimates $\E_\nu[V_x]=\sigma^2-\Var_\nu(O)$, which falls as the floor rises. An interval that adds the floor to the finite-population sampling variance covers $\mu$ in $94$ to $96\%$ of fields at every correlation length, and the certificate refuses a $10\%$ tolerance once $\kappa\sqrt{R_n}$ exceeds it, near $\xi/L=0.04$. For a network of monitors reading such a field, a tolerance below $\kappa\sqrt{R_n}$ cannot be met by adding monitors. Independent observations in time are needed instead.

\section{Floor estimation}\label{sec:estimation}

\cref{cor:floor} requires $\Var_\nu(O)$, an ensemble quantity, whereas often only one realisation is available, such as one long simulation or one observed field.

\begin{question}\label{q:single}
From a single realisation, can $R_n$ be estimated with a guarantee, or bounded above conservatively?
\end{question}

We propose \cref{alg:floor}, which is based on batch means \cite{FlyvbjergPetersen1989,Geyer1992}. For blocks of $b$ components much larger than the correlation length, $q_b=b\,\Varh_b$, with $\Varh_b$ the variance of the block means, approaches $n\Var_\nu(O)$.

\begin{algorithm}[t]
\caption{Single-realisation floor estimate (proposed)}
\label{alg:floor}
\begin{algorithmic}[1]
\Require one realisation, block sizes $\{b\}$, tolerance $\delta$, level $\alpha$
\State compute the means of non-overlapping blocks of $b$ components, and $q_b\gets b\,\Varh_b$
\State extrapolate $q_b$ to a plateau $\hat q_\infty$ as $b$ grows
\State set $\hat R_n\gets\hat q_\infty/(n\hat O^2)$, with $\hat O$ the realisation's mean, and inflate it for the block count to $\hat R_n^+$
\If{$\delta\le\kappa\sqrt{\hat R_n^+}$} refuse
\Else{} certify with $m_n^\star$ of \eqref{eq:mstar-n}, using $\hat R_n^+$
\EndIf
\end{algorithmic}
\end{algorithm}

We validated \cref{alg:floor} on the field of \cref{sec:fields}, whose floor is known exactly, using square blocks of side $s=1,2,4,\dots,L/4$ ($b=s^2$, which leaves $k=16$ blocks at the largest size), the extrapolation $\hat q_\infty=\max(q_{L/4},2q_{L/4}-q_{L/8})$ indexed by side, and inflation by $(k-1)/\chi^2_{\alpha,k-1}$.

\begin{figure}[t]
\centering
\includegraphics[width=0.55\textwidth]{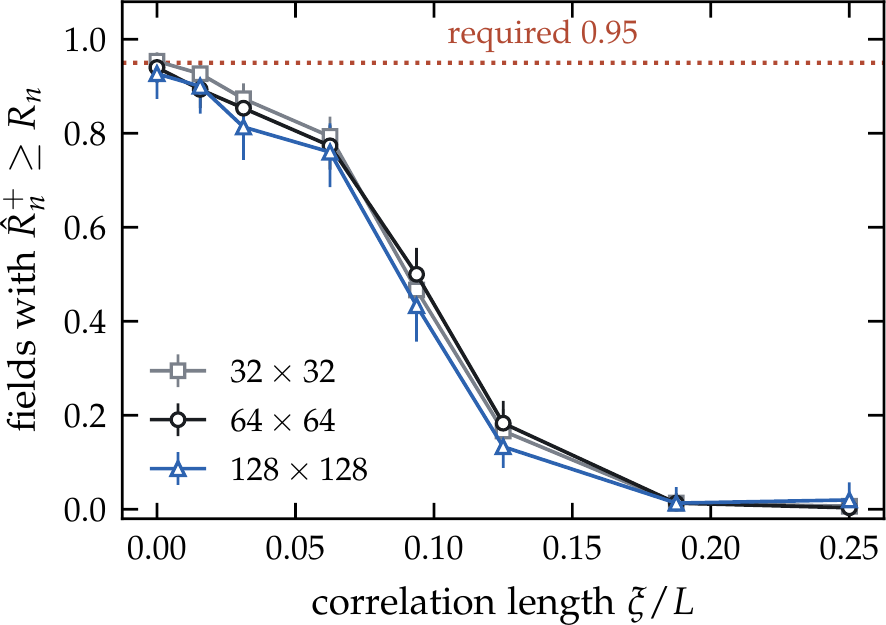}
\caption{Fraction of fields in which the bound of \cref{alg:floor} holds, against the exact floor, for three grid sizes ($300$ fields per point, $150$ at $128^2$, with Wilson 95\% intervals). The bound holds near the required 95\% only for independent sites and fails for most fields beyond $\xi/L\approx0.09$.}
\label{fig:alg}
\end{figure}

\cref{fig:alg} shows the result. For independent sites, the bound holds in $93$ to $95\%$ of fields. Its coverage is $81$ to $87\%$ at $\xi/L\approx0.03$, $76$ to $79\%$ at $\xi/L=0.0625$, below one half near $\xi/L=0.09$ and about $1\%$ for $\xi/L\ge0.19$, where the median bound is between a fifth and a twelfth of the true floor.

At $\xi/L=0.25$ about $9\%$ of fields would be certified at $\delta=10\%$, although the true floor forbids it, while at $\xi/L\approx0.03$ about two thirds of the certifiable fields are refused. With $\xi=2$ fixed and grids from $32^2$ to $256^2$, $\hat R_n^+/R_n$ settles near $2$, so the bound shrinks at the ensemble rate when $\chi_n=\Theta(1)$. The curves for the three grids coincide when plotted against $\xi/L$, so the failure is governed by $\xi/L$, and the algorithm as instantiated is not guaranteed once $\xi/L$ exceeds a few per cent.

The failure is not specific to this algorithm. Some assumption beyond stationarity is necessary, and in practice it is an assumption on the decay of correlations.

\begin{proposition}\label{prop:nonergodic}
Let $X$ be a stationary field and $Z$ an independent random constant with variance $\omega^2$, and set $X'=X+Z$. Then $X'$ is stationary with $\Var(O')=\Var(O)+\omega^2$, while every function of the deviations $X'(x)-\bar X'$ has the same law as for $X$. Hence no almost surely finite upper bound on $\Var_\nu(O)$ computed from those deviations, such as $\hat q_\infty/n$, holds with fixed probability for every stationary field.
\end{proposition}

\begin{proof}
Adding $Z$ leaves every deviation from the realisation's mean unchanged and adds $\omega^2$ to the variance of the mean. As $\omega\to\infty$, the true floor exceeds any bound whose law does not depend on $\omega$.
\end{proof}

The field $X'$ is stationary but not ergodic. This is the formal reason why Markov chain Monte Carlo practice compares several chains \cite{GelmanRubin1992}.

\section{Discussion}\label{sec:discussion}

The certificate compares the mean of one observable against a one-parameter family of deviations, so it is necessary but not sufficient for equilibrium. \cref{thm:halfwindow} measures the gap for the shuffle. Built on the average displacement, the certificate loses sight after about $\frac n4\ln m$ steps, and the count of fixed cards in the same sample keeps it for about twice as many. The criterion \eqref{eq:cert} assumes an approximately normal estimate whose variance is known and equal under both hypotheses. This fails at the level of the floor for $|M|$, whose law at $T_c$ is not normal, and for the fixed-point count, which is asymptotically Poisson. The Ising results are numerical, on lattices up to $128\times128$, and carry corrections to scaling. Only the mean-field counterpart is proved. Uniform random access to components is assumed throughout, the field results concern one stationary model on a torus, and \cref{alg:floor} has been tested on one system only. The effect of critical autocorrelation on the algorithm has not been examined. The Ising energy at $T_c$, whose floor is known exactly, is the natural next test.

\section{Conclusion}\label{sec:conclusion}

For the configuration in hand, a conjugate Bayesian treatment of the random transposition shuffle gives a relative uncertainty $1/\sqrt{2m}$ and a minimum informative sample $m^\star\approx z^2/(2\varepsilon^2)$, independent of $n$. For the equilibrium value, the law of total variance adds a floor $R_n$ below which no tolerance can be certified at any sample size, and whether a partial view suffices is decided by how $R_n$ scales with $n$. For the shuffle, the floor vanishes like $2/(5n)$. Before equilibrium, the mean displacement is exactly $\bigl(1-(1-2/n)^t\bigr)\AD_\infty$, and a certificate built on it loses sight after about $\frac n4\ln m$ transpositions, half as long as the count of fixed cards in the same sample. At the Ising critical point the floor of the energy vanishes while that of the magnetisation remains near $0.072$. The main outcome is a criterion set by $R_n$. Wherever the floor is known from an exact solution or from finite-size scaling, it decides before any sampling whether a target precision can be reached by observing more of one configuration or whether independent replicates are required, as for critical order parameters in Monte Carlo simulation and for correlated fields monitored by sensor networks. The central open problem is to determine the decay condition under which the floor can be bounded from a single realisation.

\section*{Acknowledgements}

The research was initially undertaken at the Research Science Initiative (RSI) India at the Indian Institute of Science, Bengaluru, under the direction of Professor Arvind Ayyer of the Department of Mathematics, where it won the Outstanding Written Presentation Award. The work has since been extended independently into the equilibrium floor, Ising criticality, and correlated random fields.

\section*{Competing Interests}
The author has no competing interests to declare that are relevant to the content of this article.

\section*{Data Availability}
All data supporting the findings of this study were generated by simulation and exact computation. No external datasets were used. The code and the generated data files are available at \url{https://github.com/mihikasinghania14-max/inference-modelling}.

\newcommand{\etalchar}[1]{$^{#1}$}

\end{document}